\documentclass[twocolumn,aps,superscriptaddress,floatfix]{revtex4-1}
\usepackage{graphicx}
\usepackage{amsmath}
\usepackage{natbib}

\begin{document}

\title{Thickness dependence of spin-orbit torques generated by WTe$_2$}
\author{David MacNeill}
\affiliation{Department of Physics, Cornell University, Ithaca, NY 14853, USA}
\author{Gregory M. Stiehl}
\affiliation{Department of Physics, Cornell University, Ithaca, NY 14853, USA}
\author{Marcos H. D. Guimar\~{a}es}
\affiliation{Department of Physics, Cornell University, Ithaca, NY 14853, USA}
\affiliation{Kavli Institute at Cornell, Cornell University, Ithaca, NY 14853, USA}
\author{Neal D. Reynolds}
\affiliation{Department of Physics, Cornell University, Ithaca, NY 14853, USA}

\author{Robert A. Buhrman}
\affiliation{Department of Applied and Engineering Physics, Cornell University, Ithaca, NY 14853, USA}
\author{Daniel C. Ralph}
\affiliation{Department of Physics, Cornell University, Ithaca, NY 14853, USA}
\affiliation{Kavli Institute at Cornell, Cornell University, Ithaca, NY 14853, USA}
\date{{\small \today}}

\begin{abstract}

We study current-induced torques in WTe$_2$/permalloy bilayers as a function of WTe$_2$ thickness. We measure the torques using both second-harmonic Hall and spin-torque ferromagnetic resonance measurements for samples with WTe$_2$ thicknesses that span from 16 nm down to a single monolayer. We confirm the existence of an out-of-plane antidamping torque, and show directly that the sign of this torque component is reversed across a monolayer step in the WTe$_2$. The magnitude of the out-of-plane antidamping torque depends only weakly on WTe$_2$ thickness, such that even a single-monolayer WTe$_2$ device provides a strong torque that is comparable to much thicker samples. In contrast, the out-of-plane field-like torque has a significant dependence on the WTe$_2$ thickness.  We demonstrate that this field-like component originates predominantly from the Oersted field, thereby correcting a previous inference drawn by our group based on a more limited set of samples.

\end{abstract}
\maketitle

Current-induced torques in materials with strong spin-orbit coupling provide an attractive approach for efficiently manipulating nanomagnets \cite{current2012Brataas}. Spin-orbit torques are most commonly studied in polycrystalline ferromagnet/heavy-metal bilayers\cite{PhysRevLett.101.036601,pi2010tilting, miron2010currentdriven, PhysRevLett.106.036601,Miron2011Perpendicular,PhysRevLett.109.096602, Liu555, Pai2012spintransfer}, but several groups have also investigated crystalline spin-orbit materials \cite{chernyshov2009evidence, Endo2010currentinduced, fang2011spinorbit, kurebayashi2014anantidamping, Wadley587, skinner2015complementary, roomtempNiMnSb2016,  fang2011spinorbit,controlWTe22016}. Using non-centrosymmetric crystals, researchers have demonstrated spin-orbit torques within a single ferromagnetic layer \cite{ kurebayashi2014anantidamping,roomtempNiMnSb2016, chernyshov2009evidence, Endo2010currentinduced} and electrical switching of an antiferromagnet \cite{Wadley587}. For some low-symmetry crystal structures, it is possible to generate out-of-plane polarized spin injection in response to an in-plane applied current \cite{controlWTe22016}. This is an important capability for applications. Out-of-plane spin injection could enable efficient antidamping switching of high-density magnetic memory devices with perpendicular magnetic anisotropy that is not possible with conventional spin-orbit torques \cite{controlWTe22016}.

\begin{figure}[!tbp]
\includegraphics[width=9 cm]{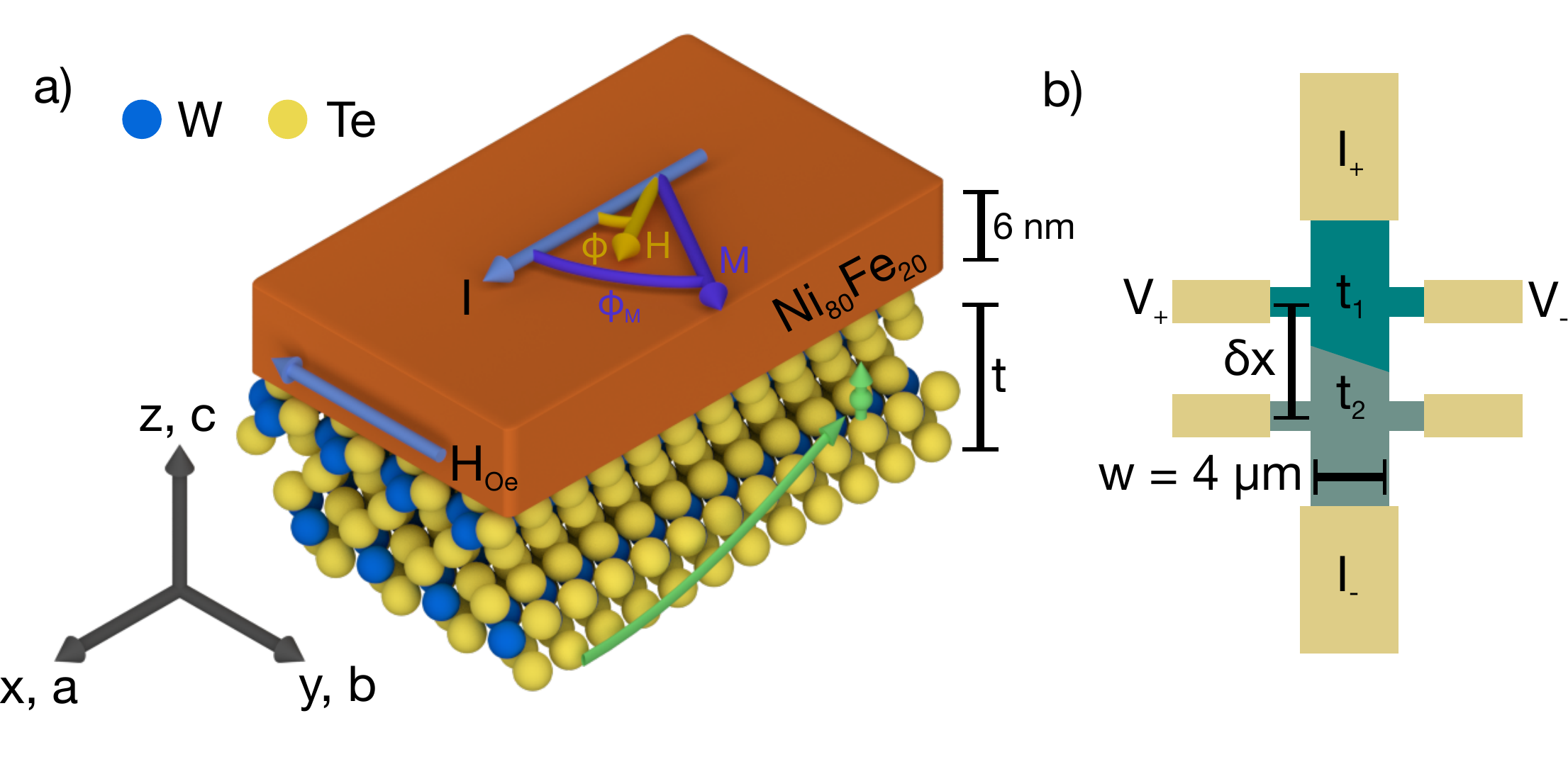}
    \caption{a) Illustration of our WTe$_2$/Py bilayers. The Py thickness is 6 nm, and the WTe$_2$ thickness, $t$, varies between devices. For all devices we study, the WTe$_2$ $c$-axis is normal to the sample plane, and the current flow direction is chosen to be approximately aligned to the WTe$_2$ $a$-axis. We carry out our measurements with the magnetic field applied at a variable angle, $\phi$ from the current flow direction. The green arrow depicts injection of out-of-plane spins into the permalloy, which can account for an out-of-plane antidamping torque. b) Illustration of the device geometry and electrical connections. For some devices, we used WTe$_2$ with mono- or bi- layer steps in the channel, allowing for multiple thickness data points from a single device. To eliminate cross talk, we keep $\delta x>4$ $\mu$m.}
    \label{fig1}
\end{figure}

Recently, our group has measured current-induced torques acting on a ferromagnetic layer (permalloy, Py = Ni$_{80}$Fe$_{20}$) deposited on single crystals of the layered material WTe$_2$ \cite{controlWTe22016}. WTe$_2$ is an intriguing choice of spin-source material, due to its strong spin-orbit coupling \cite{PhysRevB.92.125152,Pranab2016layer}, surface states \cite{wu2016observation,Huang2016spectroscopic}, high mobility \cite{ali2014large, PhysRevB.95.041410,Wang2015tuning}, and low-symmetry crystal structure \cite{brown1966thecrystal,PhysRevX.6.041069}. The crystal structure of WTe$_2$ is such that when current is applied along the WTe$_2$ $a$-axis, a spin-orbit torque consistent with transfer of spins oriented along the $z$-axis (out of the sample plane) is observed in the permalloy. The geometry is illustrated in Fig.\ \ref{fig1}. We refer to this torque as the out-of-plane antidamping torque, $\tau_{\mathrm{B}}$. As discussed in our previous work, the dependence of $\tau_{\mathrm{B}}$ on the current flow direction reflects the symmetries of the WTe$_2$ surface in a detailed way.

While the existence of $\tau_{\mathrm{B}}$ is consistent with symmetry constraints, its microscopic origin is not understood. Even the conventional current-induced torques in the WTe$_2$/Py system (an in-plane antidamping torque, $\tau_{\mathrm{S}}$, and an out-of-plane field-like torque, $\tau_{\mathrm{A}}$) have not yet been assigned concrete mechanisms. Known mechanisms such as the Rashba-Edelstein effect (REE) \cite{chernyshov2009evidence, EDELSTEIN1990233}  and the spin Hall effect (SHE) \cite{DYAKONOV1971459, PhysRevLett.83.1834} have distinct thickness dependencies once the layer thickness is comparable to the spin diffusion length. For this reason, varying the spin-source thickness can provide clues as to the origin of current-induced torques \cite{PhysRevLett.106.036601,reviewandanalysis, spintocharge2013, thicknessdepJkim2013,PhysRevLett.116.126601}. 

 \begin{figure*}[!tbp]
\includegraphics[width=17cm]{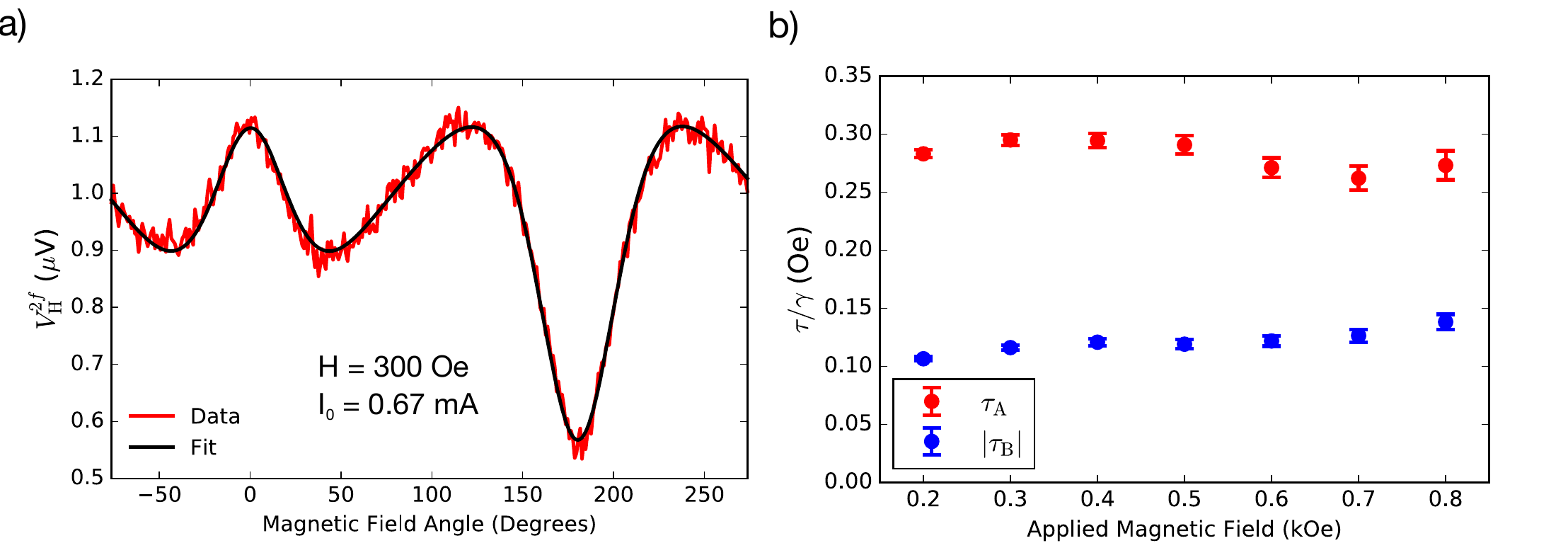}
    \caption{a) Second-harmonic Hall voltage for a WTe$_2$ (5.6 nm)/Py (6 nm) bilayer as a function of the in-plane angle of the applied magnetic field (the magnitude of the applied field is 300 Oe). The red curve represents  measured data, and the black line is a fit to Eq.\ \ref{SHfit}. The lack of odd symmetry under $\phi \rightarrow \phi+180^{\circ}$ indicates the presence of an out-of-plane antidamping toque, $\tau_{\mathrm{B}}$. b) Dependence of the measured out-of-plane field-like ($\tau_{\mathrm{A}}$, red circles) and out-of-plane antidamping torque ($\tau_{\mathrm{B}}$, blue circles) on the magnitude of applied magnetic field. The negligible field dependence is evidence that the signals arise from current-induced torques. } 
    \label{fig2}
\end{figure*}

Here, we report measurements of current-induced torques in WTe$_2$/Py bilayers for a wide range of WTe$_2$ thicknesses, down to the previously-unexplored monolayer limit.  We employ second-harmonic Hall \cite{Hwan2010Tilting, PhysRevB.89.144425} and spin-torque ferromagnetic resonance (ST-FMR) \cite{PhysRevLett.106.036601, fang2011spinorbit} measurements as complementary techniques for studying current-induced torques, and report good agreement between the two. We find that the magnitude of the out-of-plane antidamping torque component  $|\tau_{\mathrm{B}}|$ depends only weakly on the WTe$_2$ thickness $t$ for $t>4$ nm, and remains significant even for thinner samples all the way to the monolayer (0.7 nm) limit for WTe$_2$. We also demonstrate by direct measurements that the sign of $\tau_{\mathrm{B}}$ reverses across a monolayer step.  In contrast to a conclusion we made previously based on a much smaller data set \cite{controlWTe22016}, we find that the out-of-plane field-like torque varies as a function of WTe$_2$ thickness with a form in quantitative agreement with a dominant contribution from the current-induced Oersted field.

Our WTe$_2$/Py stack is shown in Fig.\ \ref{fig1}a. To prepare the stack, we take a commercially-available WTe$_2$ crystal (from HQgraphene), and exfoliate it onto a high-resistivity Si/SiO$_2$ wafer using Scotch tape. The final step of exfoliation, where the tape is removed from the substrate to cleave the WTe$_2$ crystals, is carried out in the load-lock chamber of our sputter system. The pressure at this step is well below $1\times 10^{-5}$ Torr. This preparation differs from our previous work (Ref.\ \cite{controlWTe22016}), where the samples were exfoliated in flowing nitrogen after purging the load-lock. The samples are then moved to the process chamber without breaking vacuum, where we deposit 6 nm of Py by glancing angle ($\sim 5^{\circ}$) sputtering and 2 nm of Al to prevent oxidation of the ferromagnet. The Py moment lies in the sample plane. Before further processing, the topography and thickness of the chosen flakes are characterized by atomic force microscopy (AFM). The films are patterned into Hall bars using e-beam lithography and argon ion-milling (where we use SiO$_2$ as the etch mask). The current-flow direction is chosen to lie along the direction of long straight edges in the cleaved WTe$_2$ flakes, which typically corresponds to the WTe$_2$ $a$-axis. The angle between the current flow direction and the $a$-axis is later checked using planar Hall effect measurements on the completed devices (see below). This angle was always less than 20$^{\circ}$ and typically less than 5$^{\circ}$. Contact pads of 5 nm Ti/75 nm Pt are also defined using e-beam lithography and sputtering.

We will first discuss second-harmonic Hall measurements of the spin-orbit torques.  Second-harmonic Hall measurements allow for a precise calibration of the current flowing in the device (more easily than, {\it e.g.}, ST-FMR) and therefore provide a convenient method for making an accurate comparison  between devices.  When the equilibrium magnetization is in the sample plane, this technique is most easily used for measuring out-of-plane torques because in this geometry the signals for in-plane torques must be disentangled from an artifact due to the anomalous Nernst effect \cite{PhysRevB.90.224427}. Our Hall bar design is shown in Fig.\ \ref{fig1}b. We keep the width of the channel ($w=4$ $\mu$m) and the voltage probes (1.5 $\mu$m) consistent across all devices. This helps prevent artifacts in the thickness series due to changes in the current distribution. For the second-harmonic Hall measurements, we apply a voltage of 400 mV RMS at 1.317 kHz to the device and a series resistor, and measure the first- and second-harmonic Hall voltages simultaneously. We calibrate the current flowing through the device by measuring the voltage across the series resistor. For some of our devices we placed multiple Hall contact pairs (up to three) on the same device, with each pair sensing regions of different WTe$_2$ thickness. Since the transverse voltages are expected to decay as $e^{-\pi\delta x/w}$  (see Fig.\ \ref{fig1}b) \cite{PhysRevB.79.035304}, we placed the contacts at least 4 $\mu$m apart to avoid cross-talk. This allows for direct thickness comparisons within the same device.

The Hall resistance of a WTe$_2$/Py bilayer can be modeled as $R_{\mathrm{H}}=R_{\mathrm{PHE}}\sin(2\phi_{\mathrm{M}})\sin^2(\theta_{\mathrm{M}})+R_{\mathrm{AHE}}\cos(\theta_{\mathrm{M}})$, where $\phi_{\mathrm{M}}$ is the angle between the Py moment and the current flow direction, $\theta_{\mathrm{M}}$ is the angle of the Py moment from the $z$-axis, $R_{\mathrm{PHE}}$ is the planar Hall resistance, and $R_{\mathrm{AHE}}$ is the anomalous Hall resistance. When a current $I_0 \sin(\omega t)$ is applied to the bilayer, any out-of-plane current-induced torques will rotate the moment in-plane, $\phi_{\mathrm{M}} \rightarrow \phi_{\mathrm{M}} + \delta \phi_{\mathrm{M}} \sin(\omega t)$. In-plane torques will rotate the moment out-of-plane: $\theta_{\mathrm{M}}\rightarrow \theta_{\mathrm{M}}+\delta\theta_{\mathrm{M}}  \sin(\omega t)$. The total Hall voltage due to current-induced torques is therefore $V_{\mathrm{H}}(t)=I(t)R_{\mathrm{H}}(t)=I_0R_{\mathrm{H}}\sin(\omega t)+I_0\frac{dR_{\mathrm{H}}}{d\phi_{\mathrm{M}}}\delta \phi_{\mathrm{M}}\sin^2(\omega t)+I_0\frac{dR_{\mathrm{H}}}{d\theta_{\mathrm{M}}}\delta \theta_{\mathrm{M}}\sin^2(\omega t)$. Calculating $\delta \phi_{\mathrm{M}}$ and $\delta \theta_{\mathrm{M}}$ as a function of the in-plane and out-of-plane torques, $\tau_{\phi}$ and $\tau_z$, gives the second-harmonic voltage component:

\small
\begin{equation}
\begin{aligned}
V^{2\omega}_{\mathrm{H}}\approx & I_0R_{\mathrm{PHE}}\cos(2\phi_{\mathrm{M}})\frac{\tau_{z}/\gamma}{ H+H_{\mathrm{A}}\cos(2\phi_{\mathrm{M}}-2\phi_{\mathrm{E}})}\\
&+\frac{1}{2}I_0R_{\mathrm{AHE}}\frac{\tau_{\phi}/\gamma}{ H+M_{\mathrm{s}}+H_{\mathrm{A}}\cos^2(\phi_{\mathrm{M}}-\phi_{\mathrm{E}})},
\label{SHapprox}
\end{aligned}
\end{equation}
\normalsize
where $H$ is the applied field magnitude, $M_{\mathrm{s}}$ is the effective magnetization, $H_{\mathrm{A}}$ is the in-plane uniaxial anisotropy field, and $\phi_{\mathrm{E}}$ is the angle of the anisotropy axis relative to the current flow direction. We have previously shown that the in-plane easy-axis always lies along the WTe$_2$ $b$-axis in WTe$_2$/Py bilayers (so that $\phi_{\mathrm{E}} \approx 90^{\circ}$), with an anisotropy field strength  $H_{\mathrm{A}}\approx 20-180$ Oe.  To obtain Eq.\ \ref{SHapprox}, we approximate $ \delta \phi_{\mathrm{M}}/ \delta \tau_z$ and $ \delta \theta_{\mathrm{M}}/ \delta \tau_{\mathrm{\phi}}$ at first order in $H_{\mathrm{A}}/H$. At this order, $\phi_{\mathrm{M}}=\phi-\frac{H_{\mathrm{A}}}{2H}\sin(2\phi-2\phi_{\mathrm{E}})$, where $\phi$ is the angle of the applied field from the current flow direction. The expression for $\phi_{\mathrm{M}}$ also allows $H_{\mathrm{A}}$ and $\phi_{\mathrm{E}}$ to be determined from the dependence of the first-harmonic planar Hall voltage on the angle of an applied magnetic field (see Appendix \ref{app_determination}). The results of this determination are given in Table \ref{datatab}, showing that the WTe$_2$ $a$-axis was always less than 20$^{\circ}$ from the current-flow direction.

To complete our model, we note that torques from the Oersted field and ordinary SHE will be proportional to $\hat{m}\times \hat{y}$ and $\hat{m}\times(\hat{m}\times \hat{y})$ respectively. Then $\tau_{z, \mathrm{ Oe}}(\phi_{\mathrm{M}})=\tau_{\mathrm{A}}\cos(\phi_{\mathrm{M}})$ and $\tau_{\phi, \mathrm{ SHE}}(\phi_{\mathrm{M}})=\tau_{\mathrm{S}}\cos(\phi_{\mathrm{M}})$. When a magnet absorbs out-of-plane spins the resulting torque is $\propto \hat{m}\times (\hat{m}\times \hat{z})$ \cite{Ralph20081190}, so that the out-of-plane antidamping torque gives an angle-independent contribution, $\tau_{z}(\phi_{\mathrm{M}})=\tau_{\mathrm{B}}$, for an in-plane magnetic moment. For our fits, we also add an angle-independent voltage offset, $C$, and a term $\propto \cos(\phi_{\mathrm{M}})$ to account for the anomalous Nernst effect resulting from an out-of-plane thermal gradient\cite{PhysRevB.90.224427}. The resulting model for the field and angle dependence of our second-harmonic Hall data is: 
\small
\begin{equation}
\begin{aligned}
V^{2\omega}_{\mathrm{H}}= & I_0R_{\mathrm{PHE}}\cos(2\phi_{\mathrm{M}})\frac{\left[ \tau_{\mathrm{A}}\cos(\phi_{\mathrm{M}})+\tau_{\mathrm{B}}\right]/\gamma}{ H+H_{\mathrm{A}}\cos(2\phi_{\mathrm{M-E}})}\\
                                                  & +\frac{1}{2}I_0R_{\mathrm{AHE}}\frac{\tau_{\mathrm{S}}\cos(\phi_{\mathrm{M}})/\gamma}{ H+M_{\mathrm{s}}+H_{\mathrm{A}}\cos^2(\phi_{\mathrm{M}}-\phi_{\mathrm{E}})}\\
                                                  &+V_{\mathrm{ANE}}\cos(\phi_{\mathrm{M}})+C\ \label{SHfit}
\end{aligned}
\end{equation}
\normalsize
where $V_{\mathrm{ANE}}$ is the anomalous Nernst voltage. In our system $H_{\mathrm{A}}\ll M_{\mathrm{s}}$, which means the anomalous Nernst effect and the in-plane torques give second-harmonic Hall voltages with indistinguishable $\phi$ dependence. We fit them with a single term $\propto \cos(\phi_{\mathrm{M}})$. There are six other fit parameters: $I_0R_{\mathrm{PHE}}\tau_{\mathrm{A}}$, $I_0R_{\mathrm{PHE}}\tau_{\mathrm{B}}$, $H_{\mathrm{A}}$, $\phi_{\mathrm{E}}$, $C$, and an overall angular offset not shown here which accounts for any misalignment of the device from the axes of the measurement apparatus. $I_0R_{\mathrm{PHE}}$ is determined independently using the $\phi$-dependence of the first-harmonic Hall voltage, allowing measurements of $\tau_{\mathrm{A}}$ and $\tau_{\mathrm{B}}$ from data for $V^{2\omega}_{\mathrm{H}}$ as a function of $\phi$.

 \begin{figure}[!tbp]
\includegraphics[width=9 cm]{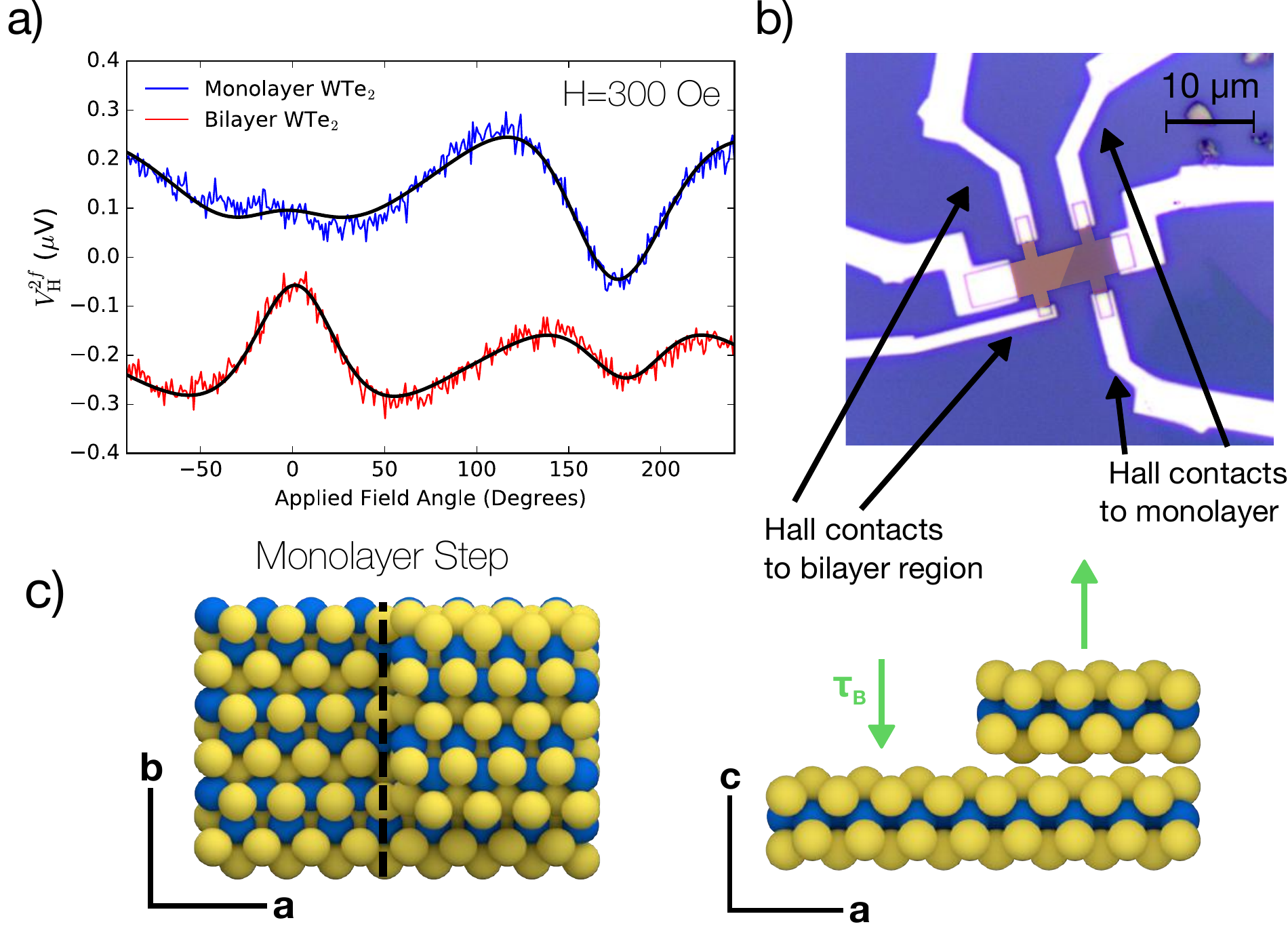}
    \caption{a) Second-harmonic Hall data for a WTe$_2$/Py device for a region of the sample with a monolayer-thick WTe$_2$ layer (top curve, blue) and for a different region of the same sample with bilayer WTe$_2$ (bottom curve, red), as a function of the angle of the applied magnetic field (defined relative to the current flow direction).  The lines are fits to Eq.\ \ref{SHfit}.  The sign reversal of $\tau_{\mathrm{B}}$ is reflected in the different angles at which the peak signals are found. A vertical offset is added to the data for ease of viewing.  b) Optical micrograph of the device measured for panel a), showing the monolayer and bilayer WTe$_2$ regions in false color. c) Schematic of the crystal structure of WTe$_2$, showing that the surface structure is rotated by 180$^{\circ}$ across a monolayer step. }
    \label{fig1_app}
\end{figure}

 \begin{figure}[!tbp]
\includegraphics[width=9 cm]{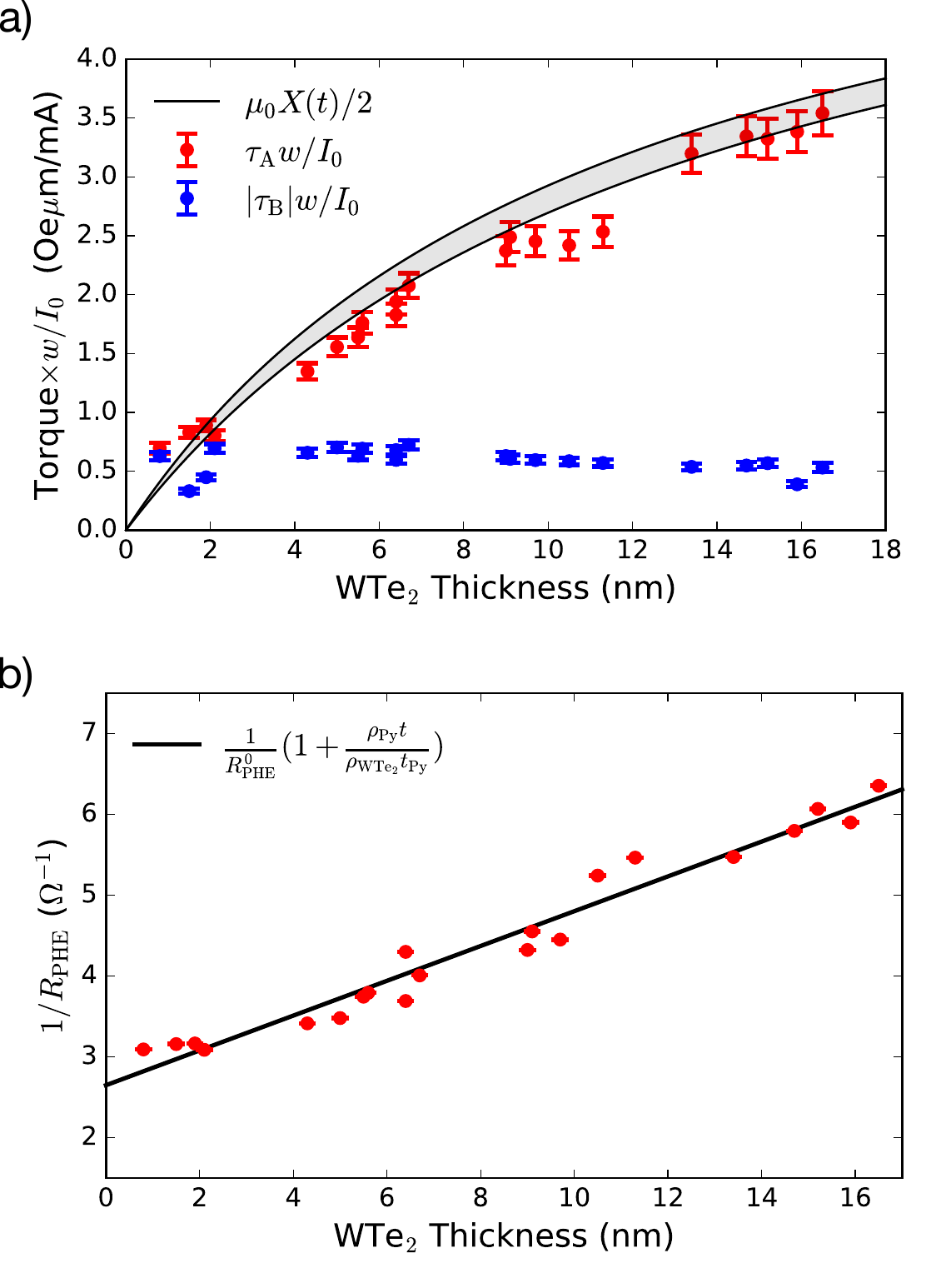}
    \caption{a) Torques normalized per unit $I_0/w$ for (red cicles) the out-of-plane field-like component $\tau_{\mathrm{A}}w/I_0$  and (blue circles) the out-of-plane antidamping component $|\tau_{\mathrm{B}}|w/I_0$, as a function of WTe$_2$ thickness, for all devices measured. The shaded region shows a $\pm$ 1$\sigma$ estimate for the torque from the magnetic field generated by the current flowing in the WTe$_2$.  b) (red circles) Dependence of the inverse of the first-harmonic planar Hall resistance on the WTe$_2$ thickness. Current shunting through the WTe$_2$ leads to a linear increase in $1/R_{\mathrm{PHE}}$ as $t$ is increased. The black line is a linear fit, which gives an estimate of the shunt factor $X(t)$ as a function of WTe$_2$ thickness.}    \label{fig3}

\end{figure}

Figure \ref{fig2}a shows $V^{2\omega}_{\mathrm{H}}(\phi)$ data from one of our WTe$_2$/Py bilayers. The WTe$_2$ is $5.6$ nm thick and the current flows along the WTe$_2$ $a$-axis ($\phi_{\mathrm{E}}\approx 90^{\circ}$). The red line shows measured data, and the black line is a fit to Eq.\ \ref{SHfit}. The existence of a non-zero value of $\tau_{\mathrm{B}}$ is apparent from the lack of $\phi\rightarrow \phi+180^{\circ}$ symmetry; in particular, the different-sized peaks at $\phi=0$ and $\phi=180^{\circ}$ relate to the cooperation $\tau_z=\tau_{\mathrm{B}}+\tau_{\mathrm{A}}$ or competition $\tau_z=\tau_{\mathrm{B}}-\tau_{\mathrm{A}}$ of the different out-of-plane torques. This asymmetry reflects the absence of rotational symmetry at the WTe$_2$ surface. Figure \ref{fig2}b shows $\tau_{\mathrm{A}}$ and $\tau_{\mathrm{B}}$ (from fits to Eq.\ \ref{SHfit}) as a function of the applied magnetic field. The extracted torques are to a good approximation independent of the magnitude of the applied field, confirming that they originate from current-induced torques.

A key prediction of our symmetry arguments in Ref.\ \cite{controlWTe22016} is that the sign of $\tau_{\mathrm{B}}$ should change across a monolayer step in WTe$_2$ thickness, if this step occurs at the Py/WTe$_2$ interface. This is because adjacent WTe$_2$ layers are related by a 180$^{\circ}$ rotation around the $c$-axis (see Fig.\ \ref{fig1_app}), and $\tau_{\mathrm{B}}$ is not two-fold symmetric -- $\tau_{\mathrm{B}}$ changes sign with a 180$^{\circ}$ rotation about the $c$-axis. In Ref.\ \cite{controlWTe22016} we presented indirect evidence for this conclusion, in which a sample whose device area spanned across a single-monolayer step in the WTe$_2$ layer exhibited a suppressed value of $\tau_{\mathrm{B}}$ due to partial cancellations of the contributions from the two crystal faces. Here we provide a direct test by fabricating devices containing multiple Hall contacts so that we can separately measure the values of $\tau_{\mathrm{B}}$ produced by different regions of the same sample separated by steps of known height (see Fig.\ \ref{fig1}). We have fabricated 6 devices with Hall contacts on either side of a monolayer step, as determined by AFM measurements showing a step height 0.7 $\pm$ 0.3 nm.  Fig.\ \ref{fig1_app} shows second-harmonic Hall data for a device where the WTe$_2$ thickness increases from a monolayer to a bilayer in the middle of the channel. For the monolayer side we found $\tau_{\mathrm{B}}/\gamma=-0.093\pm0.002$ Oe whereas for the bilayer side $\tau_{\mathrm{B}}/\gamma=0.049\pm0.002$ Oe. The out-of-plane field-like component $\tau_{\mathrm{A}}$ has the same sign on both sides of the step ($\tau_{\mathrm{A}}/\gamma=0.103\pm 0.004$ Oe and $\tau_{\mathrm{A}}/\gamma=0.123\pm 0.003$ Oe for the monolayer and bilayer respectively). In 5/6 devices with Hall contacts on opposites sides of a monolayer step, we found that $\tau_{\mathrm{B}}$ changes sign between contacts (see Appendix \ref{datatabapp}). In principle, the monolayer step we observe by AFM could be on either the top (Py/WTe$_2$) or bottom (WTe$_2$/SiO$_2$) interface of the WTe$_2$, and we do not expect that a step at the WTe$_2$/SiO$_2$ interface would affect the sign of $\tau_{\mathrm{B}}$. Therefore it is somewhat surprising that we observe sign changes in more than 50\% of samples.  It may be that the mechanics of exfoliation cause steps in the WTe$_2$ to be more likely on the top surface of the flake than the bottom. In devices with a bilayer step dividing two sets of Hall contacts, $\tau_{\mathrm{B}}$ never changes sign (3/3 devices). 

 \begin{figure}[!tbph]
\includegraphics[width=9 cm]{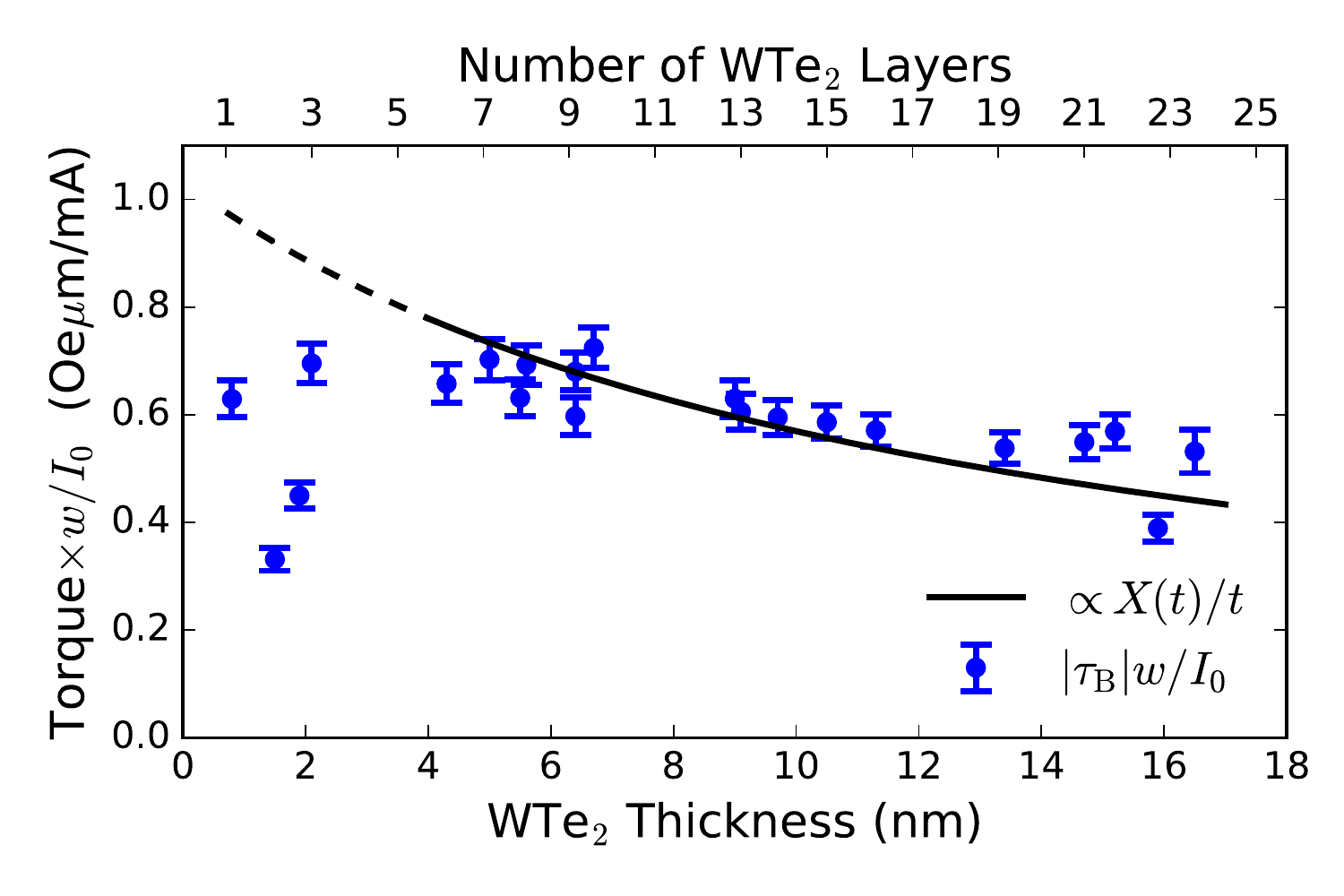}
    \caption{  $|\tau_{\mathrm{B}}|w/I_0$ as a function of WTe$_2$ thickness (blue circles), along with  a curve proportional to $X(t)/t$ as estimated from our planar Hall effect data (black solid and dashed lines). The proportionality constant is chosen to fit the data above 4 nm of WTe$_2$ thickness.}
    \label{fig4}
\end{figure}

We now turn to our thickness series over multiple devices. In total, we measured torques from 12 distinct devices, some with multiple Hall contacts per device. The resulting data are shown in Fig.\ \ref{fig3}a, where we plot $|\tau_{\mathrm{B}}|w/I_0$ (blue points) and $\tau_{\mathrm{A}}w/I_0$ (red points) as a function of WTe$_2$ thickness. The complete data set is given in Appendix \ref{datatabapp}. We normalize the torques by the current density $I_0/w$ since we can measure the current flowing in the channel more easily than the electric field.  We observe in Fig.\ \ref{fig3}a that the out-of-plane field-like torque $\tau_{\mathrm{A}}/I_0$ has a significant dependence on the WTe$_2$ thickness, increasing by a factor of over 4.8 between the monolayer sample and 16 nm, while the out-of-plane antidamping torque has a much weaker dependence.

 \begin{figure}[t]
\includegraphics[width=9 cm]{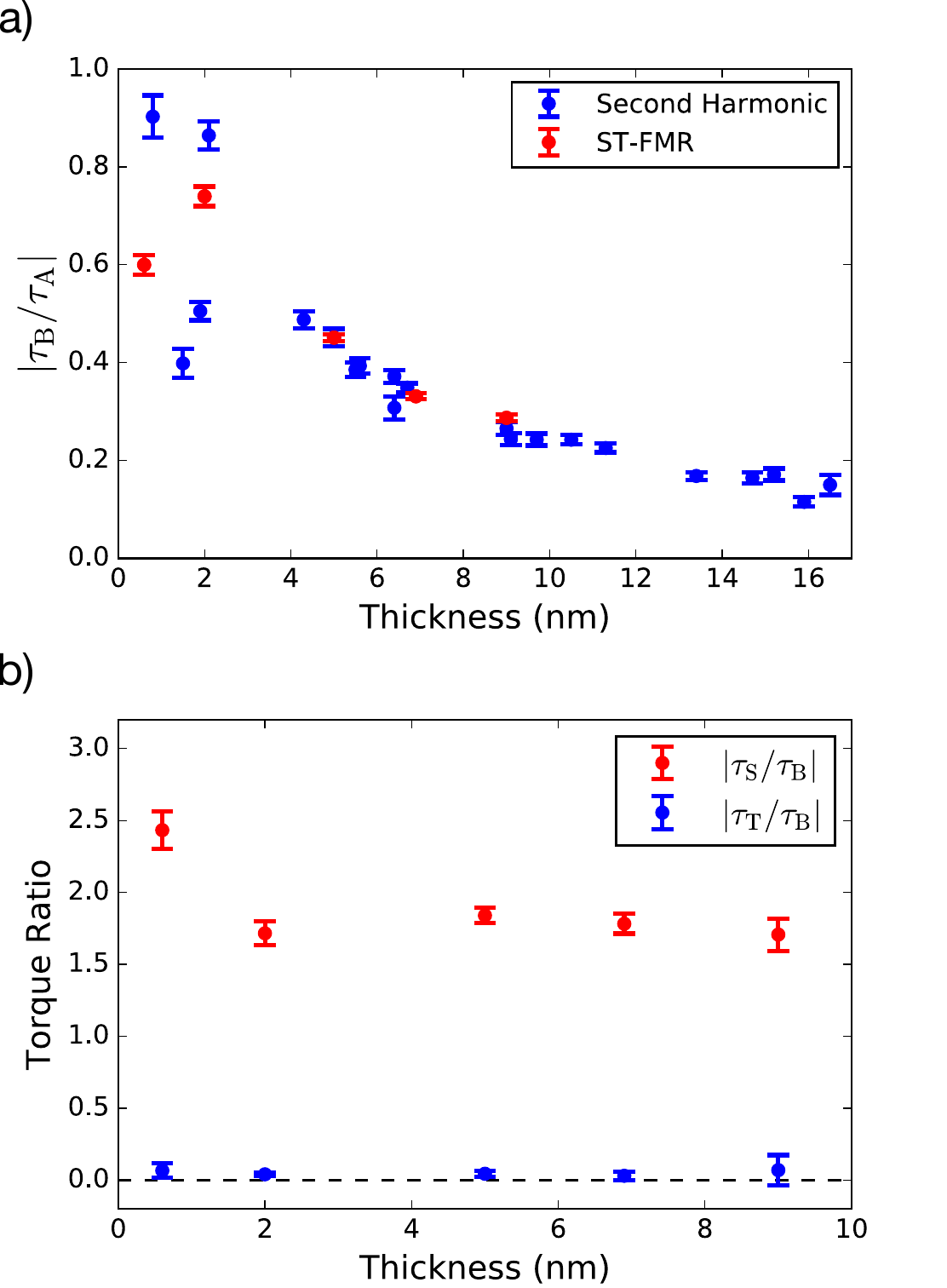}
    \caption{a) Comparison of the torque ratios $|\tau_{\mathrm{B}}/\tau_{\mathrm{A}}|$ from ST-FMR and second-harmonic Hall measurements for WTe$_2$/Py bilayers, as a function of thickness. The blue circles give $|\tau_{\mathrm{B}}/\tau_{\mathrm{A}}|$ from the second-harmonic Hall measurements, and the red circles are the values from ST-FMR. For all ST-FMR measurements, the applied frequency was 9 GHz, and for the second-harmonic measurements, the applied magnetic field was 300 Oe. b) (red circles) Ratios of the in-plane antidamping torque $\tau_{\mathrm{S}}$ to the out-of-plane antidamping torque $\tau_{\mathrm{B}}$ as a function of WTe$_2$ thickness. (blue circles) Ratios of the in-plane field-like torque $\tau_{\mathrm{T}}$ to the out-of-plane antidamping torque $\tau_{\mathrm{B}}$ as a function of WTe$_2$ thickness. The latter ratio is zero within our measurement uncertainty.}
    \label{fig2_app}
\end{figure}

In many spin-orbit torque systems (but not all \cite{mellnik2014spintransfer, skinner2015complementary, fang2011spinorbit,ciccarelli2016roomtemp}), the out-of-plane field-like torque $\tau_{\mathrm{A}}$ is dominated by a contribution from the Oersted field. The Oersted torque is related to the fraction of current flowing in the non-magnetic underlayer, $X(t)\equiv I_{\mathrm{WTe_{2}}}/I_0$, by $\tau_{\mathrm{Oe}}=\mu_0 X(t)I_0/2w$ where  $I_0= I_{\mathrm{Py}}+I_{\mathrm{WTe_{2}}}$. To determine the factor $X(t)$ for our devices, we examine the planar Hall resistance extracted from the first-harmonic Hall voltage as a function of $t$ (shown in Fig.\ \ref{fig3}b). The observed  linear dependence on WTe$_2$ thickness is consistent with a reduction in the planar Hall resistance due to shunting through the WTe$_2$, and an approximately-constant WTe$_2$ resistivity:
\begin{equation}
\frac{1}{R_{\mathrm{PHE}}}=\frac{I_{\mathrm{Py}}+I_{\mathrm{WTe_{2}}}}{V_{\mathrm{PHE}}}=\frac{1}{R^{0}_{\mathrm{PHE}}}[1+\frac{\rho_{\mathrm{Py}}t}{\rho_{\mathrm{WTe}_2}t_{\mathrm{Py}}}],
\end{equation}
where $1/R^{0}_{\mathrm{PHE}}\equiv I_{\mathrm{Py}}/ V_{\mathrm{PHE}}$ when $I_{\mathrm{Py}}=I_0$. The fit yields a normalized WTe$_2$ conductivity of  $\rho_{\mathrm{Py}}/(\rho_{\mathrm{WTe}_2}t_{\mathrm{Py}})$= 0.081 $\pm$ $0.006$ nm$^{-1}$ and a planar Hall coefficient of $R^{0}_{\mathrm{PHE}}=0.38$ $\pm$ 0.1 $\Omega$ for the Py. The normalized WTe$_2$ conductivity can be used to estimate:
\begin{equation}
X(t)\approx \frac{1}{1+\frac{\rho_{\mathrm{WTe}_2}t_{\mathrm{Py}}}{\rho_{\mathrm{Py}}t}}.
\end{equation}
The shaded black area of Fig.\ \ref{fig3}a shows the range of the expected Oersted torque (times $w/I_0$) within one standard deviation of the best-fit value for $\rho_{\mathrm{Py}}/(\rho_{\mathrm{WTe}_2}t_{\mathrm{Py}})$. The measured points for $\tau_{\mathrm{A}}w/I_0$ all fall close to this area, indicating that $\tau_{\mathrm{A}}$ is dominated by the current-generated Oersted field.  This result differs from a conclusion we drew based on a more limited data set of devices with $\phi_{a-\mathrm{I}}<20^{\circ}$ in Ref.\ \cite{controlWTe22016}. Of course, our data can not rule out additional spin-orbit contributions, which may be detected by more precise calibration of the Oersted field.

As noted above, compared to $\tau_{\mathrm{A}}w/I_0$, the out-of-plane antidamping torque $|\tau_{\mathrm{B}}|w/I_0$ displays a much weaker dependence on WTe$_2$ thickness. The form of this weaker dependence is displayed in Fig.\ \ref{fig4}, which shows a zoomed-in plot of the same data as in Fig.\ \ref{fig3}a  (blue points). For WTe$_2$ thicknesses greater than 4 nm, $|\tau_{\mathrm{B}}|w/I_0$ decreases slightly as the WTe$_2$ thickness is increased. This slight decrease is consistent with current shunting, if one assumes that $|\tau_{\mathrm{B}}|$ is proportional to the applied electric field within the device.  In this case $|\tau_{\mathrm{B}}|w/I_0$ should be proportional to $X(t)/t$. This proportionality occurs because for a given applied current $I_0$ the total electric field will decrease with increasing WTe$_2$ thickness due a decreased overall device resistance. The black line in Fig.\ \ref{fig4} shows $X(t)/t$ estimated from the PHE data of Fig.\ \ref{fig3}b, re-scaled to fit the $|\tau_{\mathrm{B}}|w/I_0$ data for WTe$_2$ thicknesses above 4 nm. This good agreement, however, tells us little about the origin of $\tau_{\mathrm{B}}$, since the total electric field in the device, the charge current density in the WTe$_2$, and the charge current density in the Py are all proportional to this factor.

For $t < 4$ nm, the measurements of $|\tau_{\mathrm{B}}|$ exhibit significantly increased scatter, but even in this regime $|\tau_{\mathrm{B}}|$ can remain large.  For the one sample with a single-monolayer WTe$_2$ that we have been able to study, we find $|\tau_{\mathrm{B}}|w/I_0 =0.63\pm 0.03$  Oe $\mu$m/mA, very comparable to the values measured for much thicker WTe$_2$ layers, and fully 65\% of the value expected simply by scaling the results from the thicker layers by the factor $X(t)/t$ (see Fig.\ \ref{fig4}).  Our observation that the torque for monolayer WTe$_2$ samples is not suppressed close to zero suggests that either the spin diffusion length in WTe$_2$ is very short, comparable to the layer spacing, or else the out-of-plane antidamping torque results from a spin current generated in the Py layer that reflects off of the WTe$_2$ surface \cite{PhysRevB.94.104419,PhysRevB.94.104420,alisha2017Observation}.
Our data for very thin WTe$_2$ layers also provides a hint that there might be an even-odd effect in the number of WTe$_2$ layers, in that $|\tau_{\mathrm{B}}|$ for a bilayer sample is the smallest for any of our devices, and in particular it is smaller than for either the monolayer sample or trilayer samples.

To confirm the results of Fig.\ \ref{fig3}a using an independent measurement technique, we also performed ST-FMR measurements using two-terminal devices fabricated from our vacuum-exfoliated WTe$_2$/Py bilayers.  The ST-FMR technique has the advantage that it can provide reliable measurements of both out-of-plane and in-plane current-induced torques, although the current calibration has greater uncertainty because this calibration must be performed using network-analyzer reflectance measurements \cite{mellnik2014spintransfer}.  For this reason, we will present our ST-FMR results in terms of ratios for the different torque components, in which case the current calibration does not enter.

For the ST-FMR samples, the WTe$_2$/Py bilayers were etched into bars and contacted in a ground-signal-ground geometry compatible with microwave probes. The device geometry and protocol for our ST-FMR measurements are detailed in Ref.\ \cite{controlWTe22016}; for the data shown here, the applied frequency was 9 GHz. Figure \ref{fig2_app}a compares the torque ratios $|\tau_{\mathrm{B}}/\tau_{\mathrm{A}}|$ measured with ST-FMR to those from second-harmonic Hall measurements as a function of WTe$_2$ thickness. The ratio $|\tau_{\mathrm{B}}/\tau_{\mathrm{A}}|$ shows good agreement with the second-harmonic Hall measurements.

Figure \ref{fig2_app}b displays the in-plane torques measured with ST-FMR. Consistent with the results in Ref.\ \cite{controlWTe22016} we measure a significant in-plane antidamping torque of the form $\tau_{\mathrm{S}}\hat{m}\times(\hat{m}\times\hat{y})$. We find that $|\tau_{\mathrm{S}}/\tau_{\mathrm{B}}|>1$ and that $|\tau_{\mathrm{S}}/\tau_{\mathrm{B}}|$ does not depend strongly on thickness. As in Ref.\ \cite{controlWTe22016}, we again note that although symmetry allows for an in-plane field-like torque of the form $\tau_{\mathrm{T}}\hat{m}\times\hat{z}$, we find that $\tau_{\mathrm{T}}=0$ within our measurement uncertainty.

In summary, we measure current-induced torques in WTe$_2$/Py bilayers as a function of WTe$_2$ thickness. We provide direct confirmation that the out-of-plane antidamping torque $\tau_{\mathrm{B}}$ changes sign across a monolayer step in the WTe$_2$, consistent with the non-symmorphic symmetries in bulk WTe$_2$.  For WTe$_2$ thicknesses $t$ greater than 4 nm, $|\tau_{\mathrm{B}}|$ decreases slowly with increasing thickness consistent with simple current shunting within the bilayer. For $t$ less then 4 nm, $|\tau_{\mathrm{B}}|$ exhibits significant device-to-device variations, which might be associated with finite size effects, interfacial charge transfer, or electronic structure changes.  Nevertheless, $\tau_{\mathrm{B}}$ remains large even for a single-monolayer of WTe$_2$.   The out-of-plane field-like torque $\tau_{\mathrm{A}}$ displays a much stronger dependence on WTe$_2$ thickness, that is quantitatively consistent with the effect of the Oersted field produced by current flowing within the WTe$_2$ layer.  This conclusion regarding the dependence of field-like torque component on WTe$_2$ thickness represents a correction of our previous report based on a more limited data set of devices with $\phi_{a-\mathrm{I}}<20^{\circ}$ \cite{controlWTe22016}.

Acknowledgements.  This work was supported by the National Science Foundation (DMR-1406333), and by the NSF MRSEC program through the Cornell Center for Materials Research (DMR-1120296). G.M.S. acknowledges support by a National Science Foundation Graduate Research Fellowship under Grant No. DGE-1144153. M.H.D.G. acknowledges support by the Netherlands Organization for Scientific Research (NWO 680-50-1311) and the Kavli Institute at Cornell for Nanoscale Science. This work made use of the Cornell Nanoscale Facility, which is supported by the NSF (ECCS-1542081) and the Cornell Center for Materials Research Shared Facilities.

\appendix

\section{Torques and magnetic anisotropy parameters for all second-harmonic Hall measurements}\label{datatabapp}
\begin{table}[h!]
\centering
\scalebox{0.7}{
    \begin{tabular}{|c| c| c| c| c| c| c| c| c| c|}
    \hline
     Device Name  & $t$ (nm)  & $L$ ($\mu$m) & $\tau_{\mathrm{A}}$ (Oe)&  $\tau_{\mathrm{B}}$ (Oe)&$H_{\mathrm{A}}$ (Oe) & $\phi_{\mathrm{E}}-90^{\circ}$ & $I_0$ ($\mu$A)  \\ 
      & $\pm$ 0.3 nm & $\pm$ 0.2 $\mu$m & &  & & (Degrees) & $\pm 0.1$ $\mu$A \\ \hline
   SH4D10S1 &  5.6 & $13 $ & 0.295(4) & -0.116(2) & 57.6(4) & 2.9(2) &670.0\\ \hline
   SH4D10S2 &  6.4 & $13$ & 0.325(7) & 0.100(3) & 61.8(5) & 2.7(2) & 670.0 \\  \hline
   SH4D7S1 &  0.8 & $12.5$ & 0.103(4) & -0.093(2) & 48(4) & -2.6(2) & 591.3\\ \hline
     SH4D7S2 &  1.5 & $12.5$ & 0.123(3) & 0.049(2) & 54.4(5) & -1.9(3)& 591.3 \\ \hline
   SH4D6S1 &  16.5 & $23.5$ & 0.473(9) & -0.071(4) & 60.9(5) & 1.7(2) & 534.3 \\ \hline
   SH4D6S2 &  15.9 & $23.5$ & 0.452(4) & 0.052(2) & 54.3(5) & 2.0(2) & 534.3\\ \hline
   SH4D6S3 &  15.2 & $23.5$ & 0.444(5) & -0.076(2) & 58.9(5) & 2.9(2)& 534.3 \\ \hline
   SH5D12S1 &  6.7 & $9.5$ & 0.410(3) & 0.143(2) & 64.7(9) & -1.7(4)& 789.7 \\ \hline
   SH5D18S1 &  2.1 & $8.5$ & 0.155(4) & -0.134(2) & 57.7(5) & 18.8(2)&770.8  \\ \hline
   SH5D26S1&  5.5 & $14.5$ & 0.249(3) & 0.096(2) & 63.1(8) & 4.2(4)&608.3   \\ \hline
   SH5D26S2 & 4.3 & $14.5$ & 0.205(3) & 0.100(2) & 60.6(2) & 4.6(4)& 608.3 \\ \hline
   SH5D25S1 &  11.3 & $10.0$ & 0.506(4) & 0.114(2) & 57.5(7) & 2.6(3) &798.6 \\ \hline
   SH5D25S2 &  10.5 & $10.0$ & 0.483(4) & 0.117(2) & 56.9(7) & 1.8(3) &798.6  \\ \hline
   SH5D29S1 &  6.4 & $17.1$ & 0.242(3) & 0.090(1) & 61.1(8) & 2.7(4)&529.4  \\ \hline
   SH5D29S2 &  5.0 & $17.1$ & 0.206(3) & 0.093(2) & 64.6(8) & 2.3(3) &529.4  \\ \hline
   SH5D28S1 &  9.7 & $17.5$ & 0.367(4) & -0.089(2) & 68.1(6) & 2.1(2) &598.4 \\ \hline
   SH5D28S2 &  9.0 & $17.5$ & 0.355(4) & 0.094(2) & 69.3(9) & 2.4(4)&598.4  \\ \hline
   SH5D32S1 &  1.7 & $7.0$ & 0.192(3) & 0.097(2) & 77.4(9) & -2.4(3) & 862.9\\ \hline
   SH5D33S1 &  13.4 & $14.0$ & 0.565(4) & -0.095(2) & 72(1) & 0.5(4) &706.7 \\ \hline
   SH5D33S2 &  14.7 & $14.0$ & 0.591(6) & -0.097(3) & 67.9(7) & 0.3(3)& 706.7 \\ \hline
   SH5D36S1 &  9.1 & $8.5$ & 0.530(6) & 0.129(3) & 96(2) & -16.1(4) &851.8\\ \hline
   
    \end{tabular}}
    \caption{ Device parameters, torques measured by the second-harmonic Hall technique (for the values of applied current $I_0$ listed in the last column), and measured magnetic anisotropy parameters for the WTe$_2$/Py bilayers analyzed in the main text. Here $\phi_{\mathrm{E}}$ is the angle of the magnetic easy-axis with respect to the current flow direction, and $H_{\mathrm{A}}$ is the anisotropy field. The number after ``S," in each device name indexes the sets of contacts on the same device. }\label{datatab}

\end{table}

\section{Determination of the magnetic easy-axis from first-harmonic Hall measurements}\label{app_determination}

\begin{figure}[t!]
\includegraphics[width=9 cm]{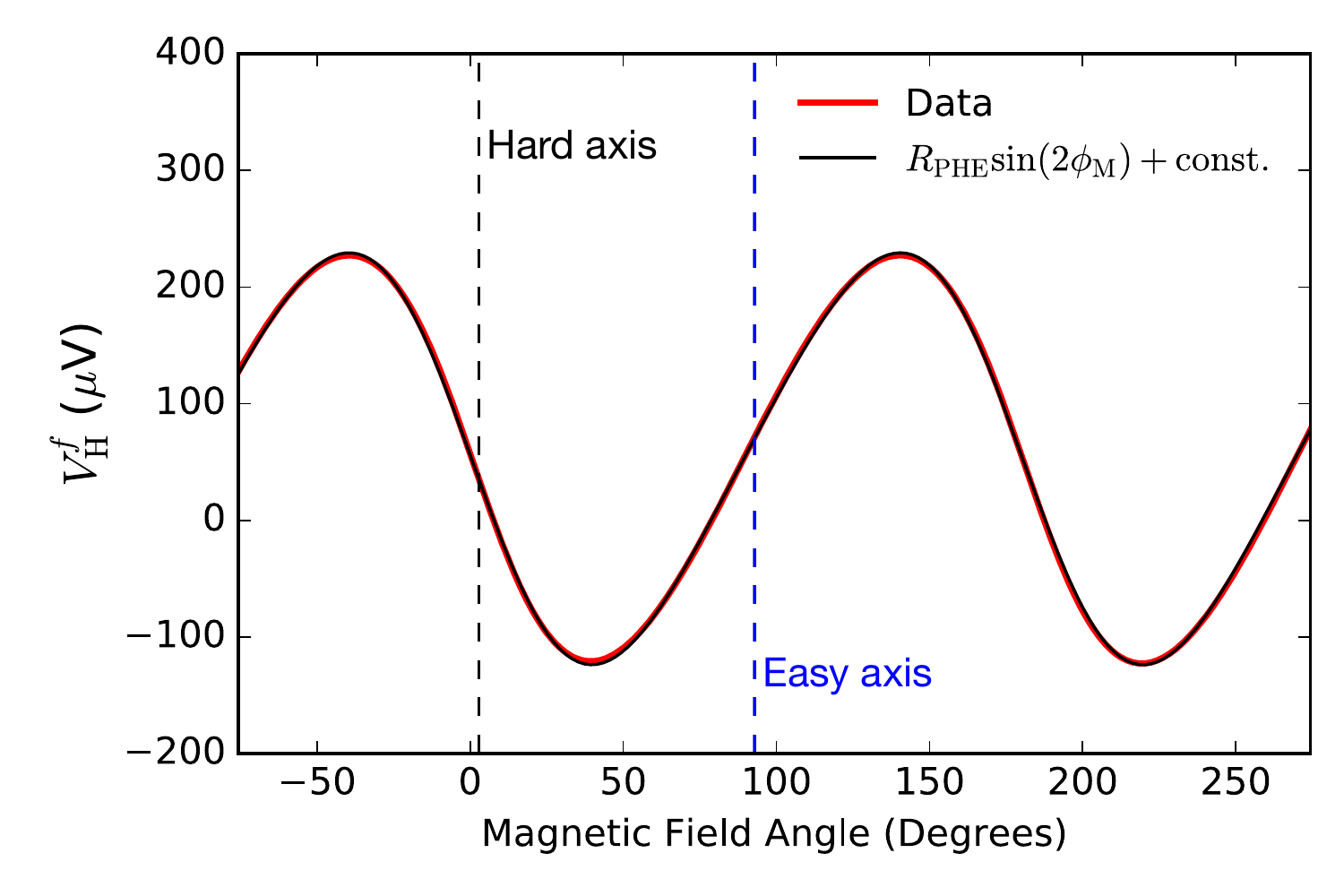}
    \caption{$V^{f}_{\mathrm{H}}$ versus $\phi$ for a WTe$_2$/Py bilayer with a 5.6 nm WTe$_2$ underlayer (red). The applied field is 300 Oe. The presence of in-plane magnetic anisotropy is apparent from the lack of symmetry around $\phi=45^{\circ}$. The solid black line is a fit assuming an in-plane uniaxial field of magnitude $H_{\mathrm{A}}$ with an easy-axis at $\phi_{\mathrm{E}}$ from the current-flow direction. The values for $\phi_{\mathrm{E}}$ and $H_{\mathrm{A}}$ determined from the fit are recorded in the ``SH4D10S1'' row of Table \ref{datatab}. The dotted black and blue lines give the estimated angles of the magnetic hard and easy axes respectively. These are equivalent to the WTe$_2$ crystal $a$ and $b$ axes.   }
    \label{firstharmfig}
\end{figure}

To confirm the alignment of the current flow direction to the WTe$_2$ $a$-axis, we use first-harmonic Hall measurements. This is possible since the WTe$_2$ $a$-axis is always along the hard direction of the in-plane uniaxial magnetic anisotropy. We previously established this fact through comparison of ST-FMR, second-harmonic Hall, and polarized Raman scattering measurements \cite{controlWTe22016}. Because of the in-plane uniaxial anisotropy, the magnetization angle of the permalloy, $\phi_{\mathrm{M}}$, will deviate slightly from the applied field angle, $\phi$. Therefore the dependence of the planar Hall effect on the applied field angle will deviate from a pure $\sin(2\phi)$ dependence, becoming:
\begin{equation}
\sin(2\phi_{\mathrm{M}})=\sin\left(2\phi-2\frac{H_{\mathrm{A}}}{2H}\sin(2\phi-2\phi_{\mathrm{E}})\right).\label{firstharmeq}
\end{equation} 
Fitting the first-harmonic Hall data to $R_{\mathrm{H}}=R_{\mathrm{PHE}}\sin(2\phi_{\mathrm{M}})$ (and a constant offset), then allows a measurement of $\phi_{\mathrm{E}}$ and $H_{\mathrm{A}}$. Data for $V^{f}_{\mathrm{H}}$ versus $\phi$, along with a fit, are given in Fig.\ \ref{firstharmfig}.

\section{Comparison between ST-FMR data from this paper and from Ref.\ \cite{controlWTe22016}}\label{app_comparison}

\begin{figure}[!tbph]
\includegraphics[width=9 cm]{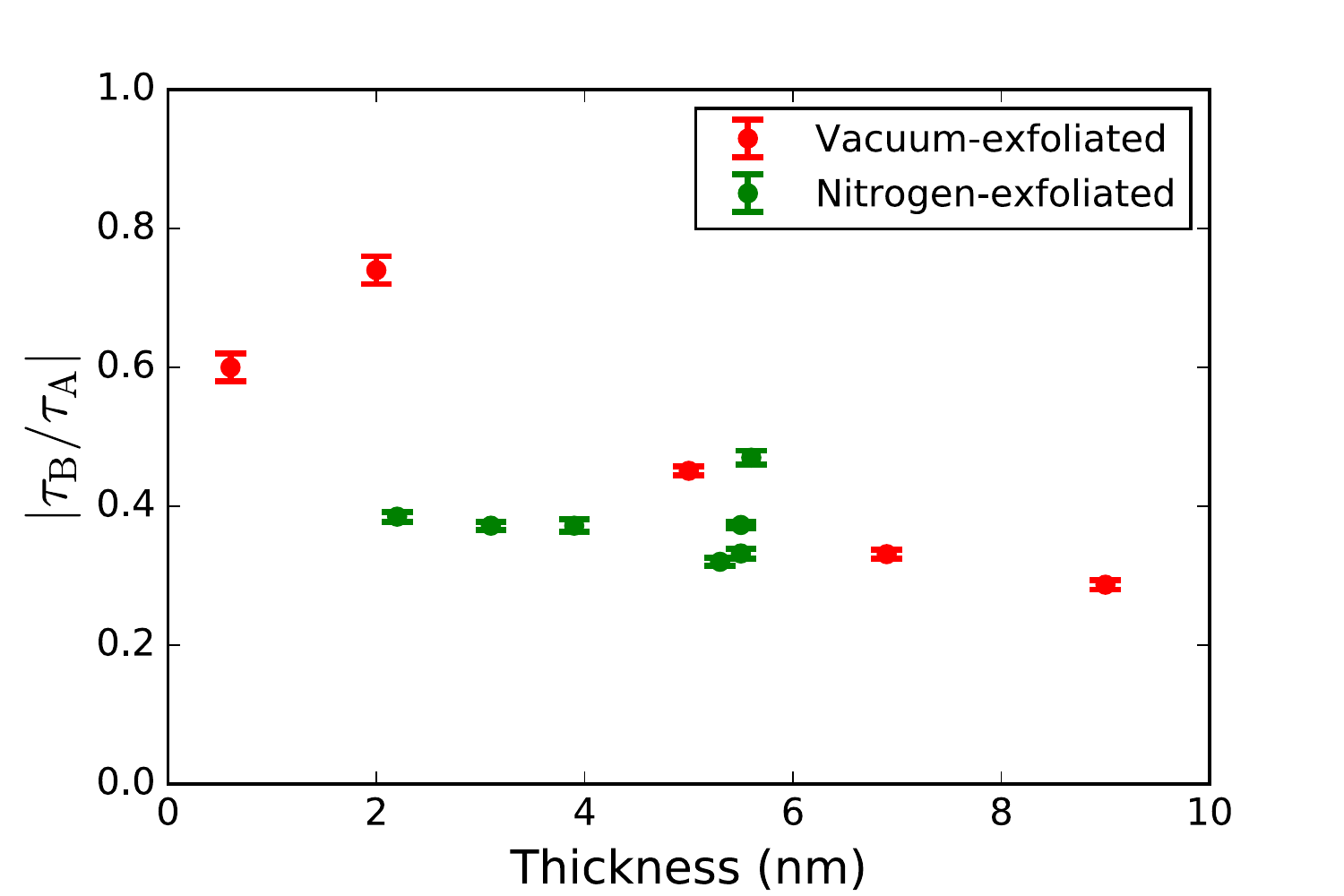}
    \caption{$|\tau_{\mathrm{B}}/\tau_{\mathrm{A}}|$ extracted from ST-FMR measurements on (green points) devices from Ref. \cite{controlWTe22016} exfoliated in flowing nitrogen and (red points) devices from this paper exfoliated in vacuum.}
    \label{fig_old_new}
\end{figure}

As discussed in the main text, for the ST-FMR data in Ref.\ \cite{controlWTe22016} we exfoliated WTe$_2$ flakes in flowing nitrogen in the load-lock chamber of our sputter system. For both the second-harmonic Hall and ST-FMR data in this paper, we exfoliated the WTe$_2$ flakes in the load-lock under vacuum better than $1\times 10^{-5}$ Torr. The ratio $|\tau_{\mathrm{B}}/\tau_{\mathrm{A}}|$ extracted via ST-FMR on the two device types is compared in Fig.\ \ref{fig_old_new}. For WTe$_2$ films around 4 nm, the vacuum exfoliated (red) and nitrogen-exfoliated (green) devices are in good agreement, whereas there is apparent disagreement for thinner flakes. We are not certain whether this apparent disagreement arises from low statistics, or from reaction of the WTe$_2$ during the nitrogen exfoliation. The effects of oxygen/water exposure on the WTe$_2$ surface merit further study.

%

\end{document}